\documentstyle[aps,preprint,floats,tighten]{revtex}
\input{psfig.tex}
\begin{document}
\draft
%\twocolumn[\hsize\textwidth\columnwidth\hsize\csname @twocolumnfalse\endcsname
\preprint{\vbox{\hbox{CU-TP-822} 
                \hbox{CAL-632}
%                \hbox{CfA-????}
                \hbox{astro-ph/9703118}
}}

\title{Getting Around Cosmic Variance}

\author{Marc Kamionkowski\footnote{kamion@phys.columbia.edu}}
\address{Department of Physics, Columbia University, 538 West
120th Street, New York, New York~~10027}
\author{Abraham Loeb\footnote{aloeb@cfa.harvard.edu}}
\address{Astronomy Department, Harvard University, Cambridge,
MA~~02138} 
\date{March 1997}
\maketitle

\begin{abstract}
Cosmic microwave background (CMB) anisotropies probe the primordial density
field at the edge of the observable Universe.  There is a limiting
precision (``cosmic variance'') with which anisotropies can determine the
amplitude of primordial mass fluctuations.  This arises because the surface
of last scatter (SLS) probes only a finite two-dimensional slice of the
Universe.  Probing other SLSs observed from different locations in the
Universe would reduce the cosmic variance.  In particular, the polarization
of CMB photons scattered by the electron gas in a cluster of galaxies
provides a measurement of the CMB quadrupole moment seen by the cluster.
Therefore, CMB polarization measurements toward many clusters would probe
the anisotropy on a variety of SLSs within the observable Universe, and
hence reduce the cosmic-variance uncertainty.

\end{abstract}

%]

\def\hatn{{\bf \hat n}}
\def\hatnprime{{\bf \hat n'}}
\def\hatnone{{\bf \hat n}_1}
\def\hatntwo{{\bf \hat n}_2}
\def\hatni{{\bf \hat n}_i}
\def\hatnj{{\bf \hat n}_j}
\def\vecx{{\bf x}}
\def\veck{{\bf k}}
\def\hatx{{\bf \hat x}}
\def\hatk{{\bf \hat k}}
\def\hatz{{\bf \hat z}}
\def\VEV#1{{\left\langle #1 \right\rangle}}
\def\cP{{\cal P}}
\def\noise{{\rm noise}}
\def\pix{{\rm pix}}
\def\map{{\rm map}}

One of the primary aims of cosmology is recovery of the primordial spectrum
of density perturbations which produced the large-scale structure in the
Universe today.  This spectrum should elucidate whether primordial
perturbations were produced by inflation, topological defects, or some
alternative mechanism.  Although galaxy surveys probe the {\it current}
mass distribution, the {\it primordial} spectrum is best probed by the
cosmic microwave background (CMB).  Large-angle CMB anisotropies from COBE
have already probed the spectrum on large distance scales.  Furthermore,
future experiments such as NASA's MAP \cite{MAP} and ESA's Planck Surveyor
\cite{PLANCK} should recover the primordial spectrum to much
smaller distance scales with unprecedented precision.  

However, there is a fundamental limit to the precision with which the CMB
can recover the amplitude of primordial fluctuations.  Theory predicts that
the primordial density field was a single realization of some random
process.  To test the theory, we would want to observe and average over a
number of realizations of the random process.  However, we have only one
Universe to observe, so there will be a sample variance, known as ``cosmic
variance'' in the average we construct.

\begin{figure}
\centerline{\psfig{figure=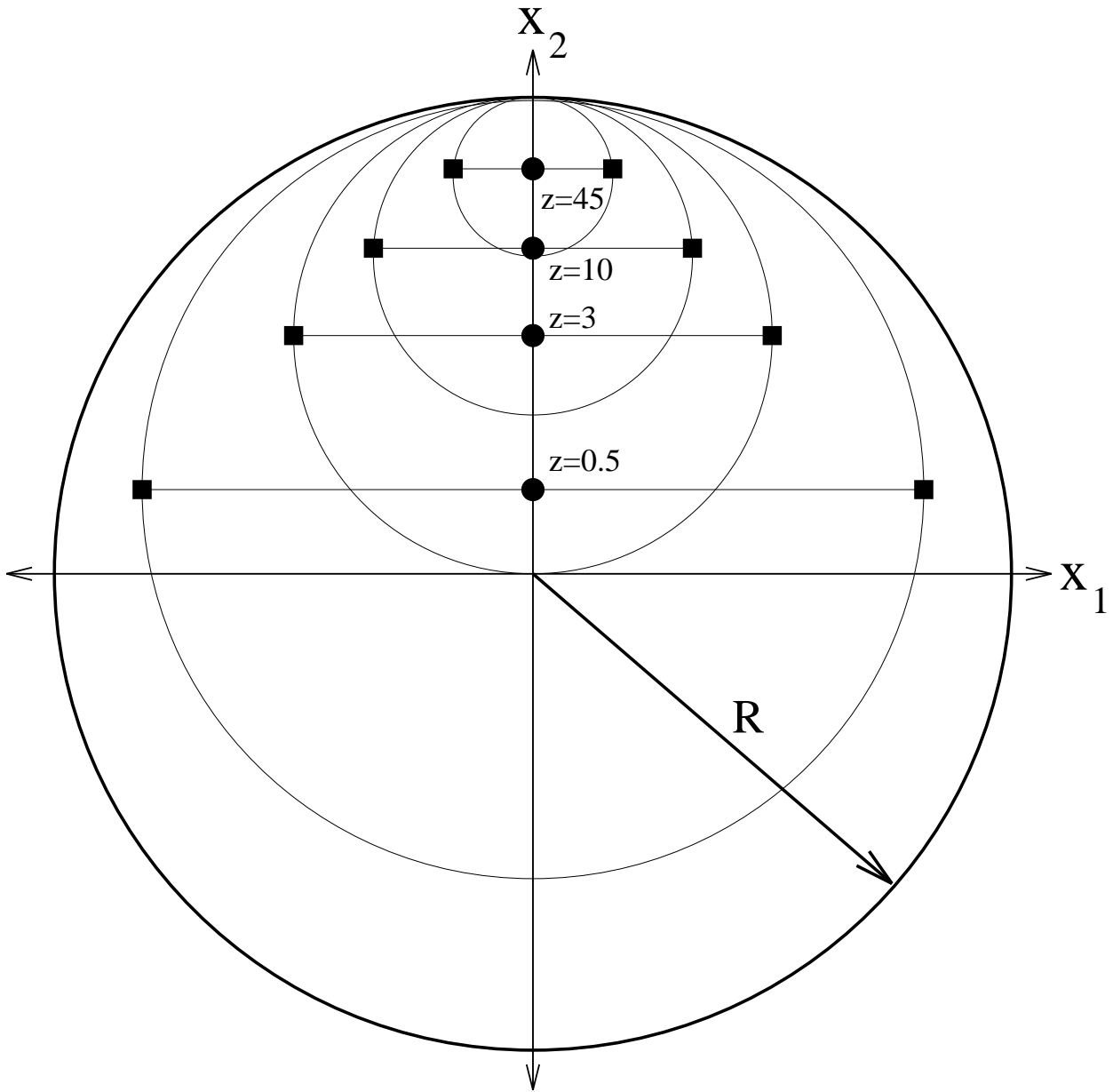,width=16.5cm}}
\bigskip
\caption{The $x_3=0$ slice of the observable Universe.  We are at
the origin and the ${\bf\hat x_3}$ direction is out of the page.  The heavy
circle is our surface of last scatter (SLS) located a comoving distance $R$
from our position at the origin.  Clusters at redshifts $z=0.5$, 3, 10, and
45 are located at the comoving positions indicated by the heavy dots along
the $+{\bf \hat x_2}$ line of sight (assuming a critical-density universe).
The circle centered on each cluster is the SLS seen by it.  The filled
squares show the primary plane of the quadrupole moment which polarizes the
radiation scattered to us.  
}
\label{thefigure}
\end{figure}

A given theory provides the three-dimensional power spectrum, $P(k)$, as a
function of wavenumber $k$.  This specifies roughly the variance in the
mass distribution over a comoving-distance scale $\lambda\sim k^{-1}$.  Now
suppose, for example, that we want to use CMB anisotropies to determine the
variance in the mass distribution averaged over spheres of comoving
diameter $\lambda$.  The CMB probes a spherical surface of last scatter
(SLS) at the edge of the observable Universe of comoving radius $R$, as
illustrated in Fig. 1.  There are only $N\sim4\pi(R/\lambda)^2$ such
volumes probed by our SLS.  Therefore, the fractional precision with which
we will be able to determine this variance in the mass distribution is
${\cal O}(N^{-1/2})$.

To be more precise, the temperature $T(\hatn)$ as a
function of direction $\hatn=(\theta,\phi)$ on the sky can be
expanded in terms of spherical harmonics,
\begin{equation}
     T(\hatn) = \sum_{lm} a_{lm} Y_{lm}(\theta,\phi),
\label{Texpansion}
\end{equation}
with
\begin{equation}
     a_{lm} = \int \, d\hatn \, T(\hatn)\, Y_{lm}^*(\hatn).
\label{Tcoefficients}
\end{equation}
These multipole coefficients $a_{lm}$ are distributed with mean
$\VEV{a_{lm}}=0$ and variance,
\begin{equation}
     \VEV{a_{l'm'}^* a_{lm}} = C_l \delta_{ll'} \delta_{mm'},
\label{alms}
\end{equation}
where the angular brackets denote an average over all realizations of the
random field.  The set of expectation values $C_l$ is the angular power
spectrum of the CMB, the projection of the three-dimensional power spectrum
$P(k)$ on the two-dimensional SLS.  Roughly speaking, $C_l$ specifies the
variance in the mass distribution on a distance scale $\pi R/l$.

To measure a given expectation value $C_l$, we would construct the $2l+1$
(for $m=-l,...,l$) $a_{lm}$ coefficients from the sky map.  The average of
the squares of these, $\widehat{C_l}=\sum_{m=-l}^l|a_{lm}|^2/(2l+1)$ would
provide our best estimate for $C_l$.
However, this average is over a sample with a finite number ($2l+1$) of
independent terms.  Therefore, the precision with which the estimator
($\widehat{C_l}$) will recover the expectation value ($C_l$) is limited.
If the distribution of density perturbations is Gaussian, the $1\sigma$
cosmic variance with which $C_l$ can be estimated is $[2/(2l+1)]^{1/2}C_l$
(see, e.g., Ref. \cite{kks}).  Although different cosmological models make
different predictions for the $C_l$'s measured by COBE (i.e., those for
$l\lesssim 15$), cosmic variance restricts our ability to discriminate
between these different models with COBE measurements.

If we could send observers to numerous distant locations in the Universe
and have them report back to us on the CMB anisotropies measured at each of
these locations, then we would have additional independent multipole
coefficients and therefore be able to overcome the cosmic-variance limit.
Although this is not practical, we {\it can} probe the anisotropy seen by
distant observers.

If one looks at the CMB through a cluster of galaxies, a fraction of the
photons have been scattered by the electron gas in the cluster (giving rise
to the Sunyaev-Zel'dovich (SZ) effect \cite{SZ}).  If the radiation incident
on the cluster has a quadrupole anisotropy in the plane normal to the line
of sight to the cluster, the scattered radiation will be linearly polarized
\cite{arthur,szquadrupole}.  Moreover, the polarization vector
will be determined by the
amplitude and orientation of the quadrupole anisotropy of the incident
radiation.  By determining the linear polarization of the CMB through a
distant cluster, we are measuring two components of the quadrupole moment
of the cluster's SLS.

Consider clusters located at the points along the $+{\bf \hat x_3}$ line of
sight indicated by the heavy dots in Fig. 1.  The sphere centered on each
cluster is the SLS observed by it.  Clearly, the SLSs of many clusters
spread throughout the observable Universe (through many different lines of
sight) would span the entire volume of the observable Universe.  The lines
perpendicular to the line of sight to each cluster indicate the plane of
the quadrupole anisotropy probed by the polarization of the scattered
radiation we observe.  The filled squares show the points on each SLS where
those quadrupole moments receive their greatest contribution.  Although
none of these squares lies closer than $R/\sqrt{2}$ {}from us (and the
closest is for a cluster at $z=3$), the volume accessible to the
squares is still $(1-2^{-3/2})\approx65\%$ of the observable volume.
Although Fig. 1 shows hypothetical clusters with $z>3$, the location of the
squares indicates that the accessible volume is equally well probed by a
sample of clusters with $z\lesssim3$. In principle, our approach could be
applied to other objects, such as galactic halos which exist at higher
redshifts.

Crudely speaking, each polarization signal measures differences in the
primordial density at points indicated by the squares in Fig. 1.
Therefore, by mapping the polarization of clusters throughout the
Universe, we can reconstruct the primordial three-dimensional density
field through most of the volume of the observable Universe in much the
same way as COBE maps the temperature on the sky from measured
temperature differences.  Therefore, polarization measurements of the CMB
through a number of distant clusters would allow us to probe a larger
volume of the observable Universe than that accessible just from our SLS.

The variance in the mass distribution on a comoving-length scale $\lambda$
can be measured with only $N\sim4\pi(R/\lambda)^2$ independent regions of
size $\lambda$ on our SLS.  However, the volume of the Universe contains
roughly $(R/\lambda)^3$ independent regions of size $\lambda$.  Therefore,
if we can map the primordial density field from cluster polarizations, then
the cosmic variance in the determination of the primordial amplitude of
density fluctuations could be reduced by up to ${\cal
O}(\sqrt{\lambda/R})$.

In addition to reducing the cosmic variance, the measured signals could
provide information complementary to that obtained from CMB anisotropy
experiments.  First, the SLSs for clusters at high redshift will be smaller
than ours, so their quadrupole moments probe smaller scales, comparable to
those probed by our higher-$l$ moments.  Therefore, by observing the
redshift dependence of the mean cluster polarization, we obtain an
independent measure of the shape of the power spectrum.  Furthermore, if
the Universe does not have a critical density, additional anisotropies will
be produced along the line of sight
\cite{kofman}.  Therefore, by comparing the redshift dependence of the
cluster polarization signal with the CMB anisotropy measurements, one could
separate the line-of-sight contribution from the anisotropy produced at the
SLS.

The signal imprinted by reionization on the CMB anisotropies could
also probe the density distribution over an extended volume in the
Universe.  However, the reionization signal is integrated over a
range of redshifts and cannot provide local information of the
type obtained from individual clusters.  Inference of the
three-dimensional density distribution from two-dimensional
power spectra requires model-dependent assumptions about the
ionization history.  Our polarization decomposition of the
line-of-sight and SLS anisotropies could independently confirm
results from studies of matter/CMB correlations \cite{mattercmb}. 

Finally, we consider the detectability of the signal.  The polarization
amplitude (in units of the CMB temperature) is expected to
be~$0.1\,\tau\,Q$, where $\tau$ is the optical depth of the cluster (as
inferred from X-ray observations) and $Q$ is the CMB quadrupole moment
\cite{arthur,szquadrupole}. Adopting $Q\approx 7\times10^{-6}$ and
a typical value of $\tau\sim 10^{-2}$ for a rich cluster, we get a
polarization signal $\sim10^{-8}$.  Is this detectable?  With current
technology, no.  However, the rate of progress in CMB measurements is
phenomenal.  A one-year dedicated experiment with a $\mu$K~$\sqrt{\rm sec}$
sensitivity could, in principle, measure the above signal for $\sim10^3$
clusters.

The quadrupole signal dominates over competing sources of polarization.
The intrinsic CMB polarization fluctuation is practically zero on the
$\sim1^\prime$ scale of a cluster core.  One could therefore
search for the special polarization pattern behind the cluster
associated with the scattered quadrupole \cite{szquadrupole}.
A peculiar velocity $v_\perp$ of the cluster transverse to the line of
sight induces effects of order $0.1(v_\perp/c)^2\tau$ or
$0.025(v_\perp/c)\tau^2$
\cite{transverse}, both of which are
much smaller than the quadrupole signal for the characteristic value of
$v_\perp/c\sim 10^{-3}$.  Complementary measurements of radial peculiar
velocities, using (the much stronger) kinematic SZ effect, can be combined
with the assumption of statistical isotropy to subtract the
transverse-velocity contribution to the polarization in a statistical way.
The small polarization signals induced by a second scattering of photons
from the thermal SZ effect, by scattering of radiation from internal radio
sources \cite{transverse}, or by gravitational effects \cite{recent}, have
different frequency or spatial distributions and could be separated from
the quadrupole signal.

In conclusion, although cluster polarization is inaccessible with current
instruments, its future implementation should give us a way to access other
SLSs {\it after} the MAP and Planck satellites tell us all there is to
learn from ours.

\bigskip

We thank U.-L. Pen and U. Seljak for useful comments.  This work was
supported at Columbia University by D.O.E. contract DEFG02-92-ER
40699, NASA ATP grant NAG5-3091, and the Alfred P. Sloan
Foundation, and at Harvard University by NASA ATP grant
NAG5-3085 and the Harvard Milton fund.

\end{document}